\newcounter{myctr}
\def\myitem{\refstepcounter{myctr}\bibfont\noindent\ifnum\themyctr>9\else\phantom{0}\fi\hangindent17pt\themyctr.\enskip}

\documentclass{ws-ijqi}
\usepackage{hyperref}
\usepackage[super,sort,compress]{cite}
\usepackage{bm}
\usepackage{amsmath}
\usepackage{latexsym}
\usepackage{amssymb}

\newcommand{\sM}{\mathsf{M}}

\def\RR{\mathbb R}
\def\CC{\mathbb C}

\def\Tr{\mathop{\rm Tr}\nolimits}

\def\bW{\bm{W}}
\def\bX{\bm{X}}
\def\bPi{\bm{\Pi}}
\def\im{{\rm\bf Im}}
\def\re{{\rm\bf Re}}

\def\Label#1{\label{#1}\ [\ #1\ ]\ }
\def\Label{\label}

\begin{document}

\catchline{}{}{}{}{}

\title{Alexander S. Holevo's Researches in Quantum Information Theory\\
in 20th Century}

\author{Masahito Hayashi}

\address{School of Data Science, The Chinese University of Hong Kong, Shenzhen, Longgang District, Shenzhen, 518172, China\\
International Quantum Academy (SIQA), Futian District, Shenzhen 518048, China \\
Graduate School of Mathematics, Nagoya University, Chikusa-ku, Nagoya 464-8602, Japan
\\
hmasahito@cuhk.edu.cn, hayashi@iqasz.cn}

\maketitle

\begin{history}
\received{Day Month Year}
\revised{Day Month Year}
\end{history}

\begin{abstract}
This paper reviews Holevo's contributions to quantum information theory
during the 20 century.
At that time, he mainly studied three topics, 
classical-quantum channel coding, 
quantum estimation with Cram\'{e}ro-Rao approach,
and quantum estimation with the group covariant approach.
This paper addresses these three topics.
\end{abstract}



\markboth{Masahito Hayashi}
{Alexander S. Holevo's Researches in Quantum Information Theory}

\section{Introduction}	
Alexander S. Holevo is undoubtedly one of the greatest researchers in 
quantum information theory.
In particular, before quantum information theory became a popular scientific area,
he led this research area as the leading person.
Nevertheless, this field could not attrach much attention at that time.
He is one of the greatest witnesses of 
the long winter experienced by the field of quantum information theory.
In 1994, the author started this area with reading his textbook \cite{Hol-Pro}:
\begin{flushleft}
A.S. Holevo, {\em Probabilistic and Statistical Aspects of Quantum Theory}. 
North-Holland, Amsterdam (1982). Originally published in Russian in 1980.
\end{flushleft}
At that time, 
we were still in the long winter era/period of quantum information theory.
In the later part of the 1990s, 
the author researched many of the findings that the author will eventually write about in the 20th century.
Therefore, the author is extremely grateful 
for Alexander S. Holevo's contributions to this area.

Looking back at his publication records, 
we see that his researches can be mainly divided into three streams.
His first stream is 
the first period on quantum information, which 
starts in the beginning of 1970s and ends in the middle of 1980s.
The above textbook summarizes the results in this period.
His second stream is his researches in 
quantum dynamical system with quantum stochastic process and semi-group,
which starts in the middle of 1980s and ends almost in the end of 20th century.
The second stream is related to the mathematical description of a quantum measurement process.
His third stream is the second period on quantum information, which 
starts in the following international conference.
\begin{flushleft}
{\em Third International Conference on Quantum Communication \& Measurement}, 
Mt. Fuji-Hakone Land, Japan, September 25-30, (1996).
\end{flushleft}
At the above conference, he knew the result for a pure-state channel coding by 
Hausladen \cite{HJSWW}.
Inspired by this result, he obtained its generalization to the case with mixed states
when he stayed in Tamagawa University after this conference \cite{HHC}.
This result was published as one of his most famous results \cite{Hol-TCap}:
\begin{flushleft}
A.S. Holevo,
``The capacity of the quantum channel with general signal states,''
{\em IEEE Transactions on Information Theory} {\bf 44} (1), 269-273 (1998).
\end{flushleft}
His third stream still continues until today. That is, he still actively studies quantum information theory.

Unfortunately, the author is not familiar with his second stream.
Since his contributions in his third stream are well known, 
this paper focuses on his first stream.
We review his major results in his first stream, which is composed of 
three topics.
The first topic is classical-quantum channel coding.
The second topic is 
quantum state estimation with the Cr\'{a}mer-Rao approach.
The third topic is 
quantum state estimation with the group covariant approach.

The remaining of this paper is organized as follows:
Section \ref{S2} reviews his results for 
classical-quantum channel coding.
Section \ref{S3} discusses his results for 
quantum state estimation with the Cr\'{a}mer Rao approach.
Section \ref{S4} explains his results for 
quantum state estimation with the group covariant approach.
In addition, these sections discuss how 
his results influenced the subsequent studies in the area of quantum information theory.
Section \ref{S5} is the conclusion.

\section{Classical-quantum channel coding}\Label{S2}
Classical-quantum (cq-) channel coding is
the subarea in quantum information theory.
\subsection{Information quantities}
Information quantities play a central role in channel coding.
Here, we introduce classical and quantum information quantities.
When a random variable $X$ is generated according to a distribution 
$P_X=\{P_X(x)\}_{x \in {\cal X}}$ on $ {\cal X} $,
its Shannon entropy $H(P_X)$ is defined as
\begin{align}
H(P_X):= -\sum_{x \in {\cal X}} P_X(x)\log P_X(x).
\end{align}
When two random variables $X,Y$ are subject to a joint distribution 
$P_{X,Y}$, the mutual information is given as
\begin{align}
I(P_{X,Y}):= H(P_X)-H(P_Y)-H(P_{X,Y}),
\end{align}
where $P_X$ and $P_Y$ are the marginal distribution of $P_{X,Y}$
with respect to $X$ and $Y$, respectively.
The mutual information has another form, by using relative entropy.
When two probability distributions $P_X$ and $Q_X$ are given,
the relative entropy is defined as
 \begin{align}
D(P_X\|Q_X):= \sum_{x \in {\cal X}} P_X(x) (\log P_X(x) -\log Q_X(x)).
\end{align}
Then, the mutual information is rewritten as
\begin{align} 
I(P_{X,Y}):=
\sum_{x \in {\cal X}}P_X(x)
D\big(P_{Y|X=x}\big\|P_Y\big),
\end{align}
where $P_{Y|X=x}$ is the conditional distribution
$P_{XY}(x,y)/P_X(x)$.

Next, we consider a quantum system ${\cal H}_1$.
Given a density matrix $\rho$ on ${\cal H}_1$,
its von Neumann entropy is given as
\begin{align}
S(\rho):= - \Tr \rho \log \rho.
\end{align}
When a density matrix $\rho_{1,2}$ on a joint system ${\cal H}_1 \otimes {\cal H}_2$,
the mutual information is given as
\begin{align}
I(\rho_{1,2}):=S(\rho_1)+S(\rho_2)+S(\rho_{1,2}),\Label{NXP}
\end{align}
where 
$\rho_{1}:= \Tr_2 \rho_{1,2}$
and $\rho_{2}:= \Tr_1 \rho_{1,2}$.
The mutual information also has another form, by using relative entropy.
When two densities matrices $\rho$ and $\sigma$ are given,
the relative entropy is defined as
 \begin{align}
D(\rho\|\sigma):= \Tr \rho (\log \rho -\log \sigma).
\end{align}
Then, the mutual information is rewritten as
\begin{align} 
I(\rho_{1,2})=
D(\rho_{1,2}\| \rho_{1}\otimes \rho_{2}).\Label{NXP2}
\end{align}

Next, we assume that the state $\rho_{1,2}$ is a classical-quantum state, 
i.e., it has the form
$\rho_{1,2}= \sum_{x \in {\cal X}}P_X(x) |x\rangle \langle x|\otimes \rho_{2|x}$,
where $P_X$ is a distribution on ${\cal X}$.
In addition, ${\cal H}_1$ is assumed to 
be spanned by an orthogonal basis
$\{|x\rangle\}_{x \in {\cal X}}$, and
$\rho_{2|x}$ is a density matrix on ${\cal H}_2$.
In this case, the formula \eqref{NXP} of the mutual information is rewritten as
\begin{align} 
I(\rho_{1,2})=S\Big(\sum_{x \in {\cal X}}P_X(x)\rho_{2|x}\Big)
-\sum_{x \in {\cal X}}P_X(x)S(\rho_{2|x}).
\end{align}
The formula \eqref{NXP2} of the mutual information is rewritten as
\begin{align} 
I(\rho_{1,2})=
\sum_{x \in {\cal X}}P_X(x)D\Big(\rho_{2|x}\Big\|\sum_{x \in {\cal X}}P_X(x)\rho_{2|x}\Big).
\end{align}

\subsection{Formulation and capacity theorem}
The problem of cq-channel coding
 is formulated as follows: \cite{Hel-Det,Hol-Tow,Hol-Est,Hol-Prob,Hel-QDet}
Consider the input alphabet ${\cal X}$ and the output quantum system
${\cal H}$, which is represented as a Hilbert space.
A cq-channel is given as $\bW:=\{W_x\}_{x \in {\cal X}}$,
where $W_x$ expresses the density matrix on ${\cal H}$
to express the output state with the input $x \in {\cal X}$.

To discuss a cq-channel, 
we discuss the mutual information 
when a random variable $X$ is generated according to distribution 
$P=\{P(x)\}_{x \in {\cal X}}$ on $ {\cal X} $.
Then, the mutual information with the input distribution $P$ is \cite{Levitin,Hol-Bound}
\begin{align} 
I(P,\bW)=
I\Big( \sum_{x \in {\cal X}}P(x) |x\rangle \langle x|\otimes W_{x}\Big).
\end{align}

The aim of cq-channel coding is the transmission of classical message via $n$ 
uses of the cq-channel $\bW $.
When we use the cq-channel $\bW $ $n$ times,
the channel is given as 
$\bW^{(n)}\{ W^{(n)}_{x^n}\}_{x^n \in {\cal X}^n}$ \cite{S-V,Hol-Cap}.
i.e., the output state with the input 
$x^n=(x_1, \ldots, x_n) \in {\cal X}^n$
is given as
\begin{align}
W^{(n)}_{x^n}:= W_{x_1} \otimes \cdots W_{x_n},
\end{align}
which is a density matrix on the $n$-fold tensor product system 
${\cal H}^{\otimes n}$.
We denote the set of messages by ${\cal M}:=
\{1, \ldots, \sM\}$.
An encoder is given as a map $\phi_n$ from ${\cal M}$ to ${\cal X}^n$.
When $\phi(m)$ is given as $(x_1, \ldots, x_n)$,
the output state is 
\begin{align}
W^{(n)}_{\phi(m)}= W_{x_1} \otimes \cdots W_{x_n}.
\end{align}
A decoder is given as a positive operator-valued measure (POVM)
$\bPi:=\{\Pi_m\}_{m \in {\cal M}}$ over the quantum system ${\cal H}^{\otimes n}$,
whose set of outcomes is given as ${\cal M}$.
That is, $\bPi$ is a resolution of the identity on ${\cal H}^{\otimes n}$,
i.e., $\Pi_m$ is a positive semi-definite matrix on ${\cal H}^{\otimes n}$,
and the relation $\sum_{m \in {\cal M}}\Pi_m=I$.
The pair of an encoder $\phi$ and a decoder $\bPi$ is called a code and is written as
$\Phi$.
The performance of a code $\Phi$ is evaluated by two parameters.
One is the size $|\Phi|$ of the code $\Phi$, which is given by $\sM$.
The other is the decoding error probability $e(\Phi)$,
which is given as
\begin{align}
e(\Phi):=1-\frac{1}{\sM}\sum_{m \in {\cal M}} \Tr W^{(n)}_{\phi(m)} \Pi_m.
\end{align}
The channel capacity is defined as
\begin{align}
C(\bW):=
\sup_{\{\Phi_n\}_n}
\Big\{ \lim_{n\to \infty} \frac{1}{n}\log |\Phi_n| \Big|
e(\Phi_n) \to 0
\Big\}.
\end{align}

The cq-channel coding theorem, i.e., the capacity theorem is formulated as
\begin{align}
C(\bW)
=\max_{P\in {\cal P}({\cal X})} I(P,\bW),\Label{IHN}
\end{align}
where
${\cal P}({\cal X})$ expresses the set of probability distributions on ${\cal X}$.
To show this result, we need to prove the following two inequalities.
\begin{align}
C(\bW)\le &\max_{P\in {\cal P}({\cal X})} I(P,\bW) \Label{IHN1}\\
C(\bW)\ge &\max_{P\in {\cal P}({\cal X})} I(P,\bW).\Label{IHN2}
\end{align}
To show \eqref{IHN}, he tackled the inequality \eqref{IHN1}, which is called 
the converse part in information theory.
He proved the inequality \eqref{IHN1} by dividing it into two steps.

In his paper \cite{Hol-Bound} in 1972, 
as the first step, 
Holevo showed
\begin{align}
I({\rm P}[P,\bW,\bPi])
\le 
I(P,\bW) \Label{YF1}
\end{align}
for any POVM $\bPi$ on ${\cal Y}$, 
where ${\rm P}[P,\bW,\bPi]$ is the joint distribution on ${\cal X} \times {\cal Y}$
defined as
\begin{align}
{\rm P}[P,\bW,\bPi](x,y):= P(x) \Tr W_x \Pi_y.
\end{align}
Nowadays, 
the inequality \eqref{YF1} can be shown by using 
the monotonicity of the quantum relative entropy \cite{Lindblad,Uhlmann}
for trace-preserving completely positive maps.
However, at that time, the above monotonicity was not known.
We can say that Holevo showed 
the monotonicity of the quantum relative entropy
in the case of the mutual information with cq-channel as \eqref{YF1}.

In his paper \cite{Hol-Cap} in 1979, as the second step, 
he showed that
\begin{align}
C(\bW) \le \lim_{n\to \infty} \frac{1}{n}
\sup_{P_n \in {\cal P}({\cal X}^n)} \sup_{\bPi_n}
I({\rm P}[P,^n,\bW^{(n)},\bPi_n]).\Label{YF2}
\end{align}
In addition, he pointed out the relation
\begin{align}
n \max_{P\in {\cal P}({\cal X})} I(P,\bW)
=
\max_{P_n \in {\cal P}({\cal X}^n)} 
 I(P_n,\bW^{(n)}).\Label{YF3}
\end{align}
In fact, the above relation can be shown by using the chain rule.
Combining three relations \eqref{YF1},
\eqref{YF2}, and \eqref{YF3}, he derived \eqref{IHN1}.

For the opposite direction \eqref{IHN2}, 
Hausladen et. al. \cite{HJSWW} showed the inequality \eqref{IHN2} when all $W_x$ 
are pure states.
Their key idea is typical subspaces \cite{Schumacher,JS}.
Inspired by this result \cite{HJSWW},
using the idea of typical subspaces,
Holevo showed the inequality \eqref{IHN2} with general mixed states $W_x$. 
Later, independently, Schumacher and Westmoreland \cite{S-W}
showed the same result.
In addition, with Burnashev, Holevo derived an exponential decay date of 
the decoding error probability
when the transmission rate is smaller than the capacity $C(\bW)$ and 
all $W_x$ are pure states \cite{B-H}.

Since the result \eqref{IHN} forms the foundation of quantum communication,
it motivated many studies in quantum information theory.
For example, Ogawa and Nagaoka \cite{O-N} showed the strong converse theorem for cq-channel coding.
To prove \eqref{IHN} via information spectrum method \cite{Han},
the reference \cite{H-N} invented a useful matrix inequality, which is often called Hayashi-Nagaoka inequality.
Bennet et al \cite{BSSV} considered cq-channel coding with entanglement-assistance, which leads to
reverse Shannon theorem \cite{BDHSW}, which is known as a new topic in Shannon theory.
Holevo \cite{Hol-Ent} provided a more elegant proof. Also using the
result \eqref{IHN}, Devetak \cite{Devetak} showed the coding theorem for sending a quantum state via a noisy quantum channel.
In this way, the result \eqref{IHN} yields various results in quantum information theory.

\section{Quantum estimation with Cr\'{a}mer Rao type bounds}\Label{S3}
Holevo made great contributions to quantum state estimation as well.
In this problem, we consider a parameterized family of density operators 
${\cal S}=\{\rho_{\theta}\}_{\theta \in \Theta}$
on a Hilbert space ${\cal H}$.
When $\Theta$ is a continuous subset of $\RR^d$,
an estimator is formulated as a POVM 
$\bPi$ on ${\cal H}$ whose data set is $\RR^d$.
Since $\RR^d$ is a continuous set,
its rigorous formulation requires measure theory.
A POVM $\Pi$ is defined as a map from the set ${\cal B}(\RR^d)$ of Borel sets to 
the set of positive semi-definite operators on ${\cal H}$.
The map $\Pi$ should satisfy the following conditions.
\begin{align}
\Pi(\emptyset)=0,\quad
\Pi(\RR^d)=I,\quad
\Pi(\cup_j B_j)= \sum_j \bPi(B_j),
\end{align}
where $B_j \in {\cal B}(\RR^d)$ are disjoint sets. 
We often consider the {\it unbiased} condition, which is formulated as
\begin{align}
\int_{\RR^d} \hat{\theta} \Tr \rho_\theta \Pi(d \hat{\theta})= \theta\Label{MAT}
\end{align}
for $\theta \in \Theta$.
Since the unbiased condition is too restrictive,
taking the Taylor expansion in \eqref{MAT} at $\theta_0\in \Theta$,
we often consider its relaxed version, the locally unbiased condition at
$\theta_0 \in \Theta$, which is formulated as
\begin{align}
\int_{\RR^d} \hat{\theta} \Tr \rho_{\theta_0} \Pi(d \hat{\theta})
&= \theta_0 \\
\frac{\partial}{\partial \theta^k}\int_{\RR^d} \hat{\theta}^l \Tr \rho_{\theta_0} \Pi(d \hat{\theta})
&= \delta_{k,l}.
\end{align}

To measure the precision of an estimator $\Pi$, 
we focus on the mean square matrix $V(\Pi)$ as
\begin{align}
V^{k,l}_\theta(\Pi):=
\int_{\RR^d} (\hat{\theta}^k-\theta^k)(\hat{\theta}^l-\theta^l)
 \Tr \rho_\theta \Pi(d \hat{\theta}).
\end{align}
When $\Pi$ is an unbiased estimator,
the mean square matrix $V_\theta(\Pi)$ coincides with the covariance matrix.
For $d=1$, we simply call $V_\theta(\Pi)$ 
the mean square.

Helstrom defined
the SLD $L_{\theta,k}$ as 
a Hermitian (self-adjoint) operator satisfying \cite{Hel-Min,Hel-QDet}
\begin{align}
L_{\theta,k} \circ \rho_\theta
= \frac{\partial \rho_\theta}{\partial \theta^k}. \Label{MXA2}
\end{align}
where $X\circ Y:= \frac{1}{2}(XY+YX)$
Then, Helstrom defined the SLD Fisher information matrix
$J_{\theta,k,l}$ as
\begin{align}
J_{\theta,k,l}:= \Tr L_{\theta,k} (L_{\theta,l} \circ \rho_\theta).
\end{align}
For a locally unbiased estimator $\Pi$ at $\theta$, 
using Schwarz inequality, 
we obtain 
the matrix inequality based on positive semi-definite matrices.
\begin{align}
V_\theta(\Pi) \ge J_\theta^{-1}.\Label{NMA}
\end{align}
In the following, we use matrix inequalities in the above sense.
For $d=1$, the following locally unbiased estimator achieves the equality
in \eqref{NMA}.
When $\Pi$ is the spectral decomposition of 
the operator $\frac{1}{J_\theta} (L_\theta+\theta)$,
the equality in \eqref{NMA} holds.

However, the equality in \eqref{NMA} does not hold in the multiple-parameter case.
This difficulty is caused by the non-commutativity of SLDs $L_{\theta,k}$.
The first attempt to tackle this problem was done by
Yuen and Lax \cite{Y-L}.
They considered complex parameters to identify the state, i.e., they assumed 
$\Theta \subset \CC^{d}$. 
They focused on the unbiased condition \eqref{MAT}
by replacing $\RR^d$ by $\CC^d$, and
considered right logarithmic derivatives based on complex 
parameters.
Their method works well for the quantum Gaussian family, whose mathematical formulation
is given in the references \cite{Hol-qua1,Hol-qua2,Hol-qua-e}\cite[Chapter 5]{Hol-Pro},
which has crucial non-commutativity related to the canonical observables
$Q$ and $P$.
But, it cannot be applied to a general real-multiple-parameter model.
At that time, several Russian researchers \cite{Belavkin,Sto} focused this topic.

To resolve this problem, for a general real-multiple-parameter model,
Holevo \cite{Hol-CR,Hol-Pro} defined
the right logarithmic derivative $L_{\theta,k}^R$ as 
\begin{align}
\rho_{\theta} L_{\theta,k}^R= \frac{\partial \rho_\theta}{\partial \theta^k}. \Label{MXA3}
\end{align}
Here, $L_{\theta,k}^R$ is not necessarily self-adjoint operator.
Holevo then defined the RLD Fisher information matrix
$ J^R_{\theta}$ as
\begin{align}
 J^R_{\theta,k,l}:=\Tr \rho L_{\theta,k}^R (L_{\theta,l}^R)^\dagger.
\end{align}
For a locally unbiased estimator $\Pi$ at $\theta$, 
using Schwarz inequality, 
we obtain
the matrix inequality 
\begin{align}
V_\theta(\Pi) \ge (J_\theta^R)^{-1}.\Label{NMA2}
\end{align}
In the one-parameter case, the inequality 
$ J_\theta^R\ge J_\theta$ holds so that
\eqref{NMA2} does not give a better bound than \eqref{NMA}.
However, in the multiple parameter case,
due to the effect of the off-diagonal part,
there are several cases where 
\eqref{NMA2} gives a better bound than \eqref{NMA}.
To clarify such a case, he introduced the D operator ${\cal D}_\theta$
on the set of self-adjoint operators as
\begin{align}
\rho_\theta \circ {\cal D}_\theta(X)= i[X,\rho_\theta].
\end{align}
When the space spanned by
$\{ L_{\theta,k} \}_k$ is invariant with respect to 
the D operator ${\cal D}_\theta$, which is called
the D invariant property,
Holevo \cite{Hol-CR,Hol-Pro} showed that 
\begin{align}
(J_\theta^R)^{-1}= J_\theta^{-1}+\frac{i}{2}J_\theta^{-1}
D_\theta J_\theta^{-1}\Label{CXI}
\end{align}
where
the antisymmetric matrix $D_\theta$ is defined as 
\begin{align}
D_{\theta,k,l}:=\Tr {\cal D}_\theta(L_{\theta,k}) 
(L_{\theta,l}\circ \rho_\theta).
\end{align}
This fact means that 
\eqref{NMA2} is a stricter matrix inequality than \eqref{NMA}
in the above case.

Furthermore, to utilize the imaginary part of the inequality \eqref{NMA2}, 
Holevo \cite{Hol-CR,Hol-Pro} considered the weighted sum of the components of the mean square error matrix $V_\theta(\Pi)$.
We choose a $d \times d$ positive semi-definite matrix $G$,
and focus on
\begin{align}
\Tr G V_\theta(\Pi).
\end{align}
In fact, since
it is impossible to simultaneously minimize all diagonal components
in $V_\theta(\Pi)$,
we need to handle their trade-off.
The above strategy enables us to discuss their trade-off.
To discuss this problem, Holevo focused on the following lemma.
\begin{lemma}{\cite[Lemma 6.6.1]{Hol-Pro},\cite[(2.9)]{Belavkin2},\cite[(8.1 1)]{Sto}}\Label{LL}
Given a Hermitian matrix $R$, we have
\begin{align}
\min_{V: \hbox{Hermitian} } \{\Tr GV | V \ge \pm R\}
=\Tr |\sqrt{G} R \sqrt{G}|.
\end{align}
The minimum holds when $V=
\sqrt{G}^{-1}|\sqrt{G} R \sqrt{G}|\sqrt{G}^{-1}$ .
\end{lemma}
This lemma can be easily shown by diagonalizing the Hermitian matrix $\sqrt{G} R \sqrt{G} $.

Combining the matrix inequality \eqref{NMA2} and Lemma \ref{LL}, 
Holevo \cite{Hol-CR,Hol-Pro} showed that
\begin{align}
\Tr G V_\theta(\Pi) \ge
\Tr G \re (J_\theta^R)^{-1}+
\Tr |\sqrt{G} \im (J_\theta^R)^{-1} \sqrt{G} |\Label{MNW7}
\end{align}
for a locally unbiased estimator $\Pi$.
Hence, the right hand side of \eqref{MNW7} is called the RLD bound.
When the D invariant property holds,
the relation \eqref{CXI} simplifies \eqref{MNW7} to
\begin{align}
\Tr G V_\theta(\Pi) \ge
\Tr G J_\theta^{-1}+\frac{1}{2}
\Tr |\sqrt{G}  J_\theta^{-1} D_\theta J_\theta^{-1}\sqrt{G} |\Label{MNW}.
\end{align}
When we employ \eqref{NMA} instead of \eqref{NMA2},
the lower bound is composed only of  the first term in \eqref{MNW}.
Also, Holevo \cite{Hol-Gau} also showed the equality in \eqref{MNW}.

However, the above approach works 
only when the D invariant property holds.
To resolve this problem, we consider a more general approach.
When $\Pi$ is a locally unbiased estimator,
we choose 
$d$ self-adjoint operators $\bX(\Pi)=(X^1(\Pi), \ldots, X^d(\Pi))$ as
\begin{align}
X^k(\Pi):= 
\int_{\RR^d} (\hat{\theta}^k-\theta^k)  \Pi(d \hat{\theta}).
\Label{MATM}
\end{align}
Then, we have the condition
\begin{align}
\Tr X^k(\Pi) \frac{\partial \rho_\theta}{\partial \theta^l}=\delta_{k,l}.\Label{BBR}
\end{align}
Also, Holevo \cite[(6.7.73)]{Hol-Pro} showed the matrix inequality
\begin{align}
V_\theta(\Pi) \ge Z(\bX(\Pi)),
\Label{NMA8}
\end{align}
where 
the Hermitian matrix $Z(\bX)$ is defined as
\begin{align}
Z^{k,l}(\bX):= &\Tr X^k X^l\rho_\theta .
\end{align}
Then, we define
\begin{align}
C_\theta(G,\bX):= \inf_{V: {\rm Symmetric}} \{   \Tr G V | V \ge Z(\bX)\}.\Label{ZSY}
\end{align}
Since the relation \eqref{NMA8} guarantees that the matrix $V_\theta(\Pi)$ satisfies the condition in \eqref{ZSY},
we find the inequality
\begin{align}
\Tr G V_\theta(\Pi) \ge
C_\theta(G,\bX(\Pi)).
\end{align}
Therefore, for a locally unbiased estimator $\Pi$,
we obtain 
\begin{align}
\Tr G V_\theta(\Pi) \ge
C_\theta^{HN}(G):=
\inf_{\bX} C_\theta(G,\bX),\Label{VBY}
\end{align}
where the above infimum is taken for 
$d$ self-adjoint operators $\bX=(X^1, \ldots, X^d)$ satisfying 
the condition \eqref{BBR}.

When the D invariant property holds,
Holevo \cite[Section 6.7]{Hol-Pro} showed that the lower bound in \eqref{VBY} equals 
the right hand side of \eqref{MNW}.
He also showed that 
the infimum in \eqref{VBY} is attained when 
the $d$ self-adjoint operators $\bX$ is given as
\begin{align}
X^k=\sum_{l=1}^d (J_\theta^{-1})^{k,l} L_{\theta,l}.
\end{align}
Using this fact, Holevo \cite{Hol-Gau,Hol-Pro} constructed an estimator to attain the lower bound 
\eqref{VBY} under the quantum Gaussian family.

Later, using Lemma \ref{LL}, Nagaoka solved the minimization \eqref{ZSY} as 
\begin{align}
C_\theta(G,\bX)=
\Tr \sqrt{G} \re Z(\bX)\sqrt{G}
+\Tr |\sqrt{G} \im Z(\bX)\sqrt{G}|.\Label{ZSY2}
\end{align}
Nagaoka \cite{Nag-New} wrote that 
Holevo introduced the lower bound $C_\theta^{HN}(G)$.
In fact, the aim of the paper \cite{Nag-New} was to present 
Nagaoka's new bound, which is the two-parameter case of the bound called 
Nagaoka-Hayashi bound in the recent paper \cite{CSLA}, as well as 
to present its advantage over the lower bound $C_\theta^{HN}(G)$.
Also, since the inequality \eqref{VBY} can be obtained with a few steps and 
a combination of 
the matrix inequality \eqref{BBR} and Lemma \ref{LL},
Nagaoka considered that the inequality \eqref{VBY} should be 
attributed to the contribution by Holevo.
Following to Nagaoka's idea, 
many subsequent studies \cite{HM08,YFG,Suzuki15,Suzuki16,YCH,Suzuki19,SYH,CSLA,Yamagata} called it the Holevo bound.
In these studies, 
one of the authors was directly suggested by Nagaoka or his collaborators
about the use of the terminology ``the Holevo bound''.
Recently, due to these studies,
other studies \cite{FPAD,AFD,TAD} followed this terminology without direct suggestion from Nagaoka or his collaborator.

However, Holevo \cite[Section 6.7]{Hol-Pro} considered the infimum \eqref{VBY}
only when the D invariant property holds.
Nagaoka \cite{Nag-New} wrote down the infimum \eqref{VBY} for the general case 
at the first time.
While Nagaoka's paper \cite{Nag-New} was written as a technical report,
the reference \cite{HM08} provided 
the full derivation of \eqref{VBY} with the form \eqref{ZSY2}
as a journal publication.
In the reference \cite[Remark 2]{HM08}, the author mentioned that
this bound was essentially introduced in \cite[Sec. VI-7]{Hol-Pro}
while the notation of \cite{HM08} is based on the reference \cite{Nag-New}.
At that time, the author was overconfident about the explanation in the reference \cite{Nag-New} for the reference \cite{Hol-Pro}.
The main issue of the reference \cite[Sec. VI-7]{Hol-Pro} is 
a statement different from \eqref{VBY}.
Since Nagaoka wrote down the infimum $C_\theta(G,\bX)$ as \eqref{ZSY2}
and the expression \eqref{ZSY2} has been widely used for the calculation of this bound,
it is suitable to call it the Holevo-Nagaoka bound 
as has been done in the reference \cite{Suzuki20}.
Indeed, some of the recent papers 
cited only the reference \cite{Hol-Pro} to discuss this bound
whereas it is hard to derive the bound \eqref{VBY} with the form \eqref{ZSY2}
only from the reference \cite{Hol-Pro}.
When a paper uses this bound, it might be better to cite paper \cite{Nag-New} (and reference \cite{HM08}) in addition to reference \cite{Hol-Pro}.

At a later time, the importance of the lower bound $C_\theta^{HN}(G)$ was revealed.
In the following, we discuss its importance in subsequent studies.
Nagaoka \cite{Nag-On} introduced the concept of 
the $n$-fold independent and identical distributed (i.i.d.) extension of 
a given state family ${\cal S}=\{\rho_\theta\}$ 
as ${\cal S}_n=\{\rho_\theta^{\otimes n}\}$. 
At that time, many strong objections to this extension 
were presented from the viewpoint of physics due to the no-cloning theorem.
Therefore, this research direction was not well accepted at that time.
The paper \cite{HM08} showed the asymptotic attainability of 
the lower bound $C_\theta^{HN}(G)$ 
under the i.i.d. extension of any submodel of the qubit system including the full model.
To show this fact, the paper \cite{HM08} showed the asymptotic approximation of 
the i.i.d. extension by the qubit system by the Gaussian state family,
which is called the local asymptotic normality.
The paper \cite{K-G} showed the local asymptotic normality of full parameter model 
in the qudit system, which implies the asymptotic attainability of 
the RLD bound 
under the i.i.d. extension of the full parameter model of the qubit system.
Finally, the papers \cite{YFG,YCH}
showed the asymptotic attainability of 
the lower bound $C_\theta^{HN}(G)$ 
under the i.i.d. extension of any submodel of the qubit system
in various settings.
In this way, the lower bound $C_\theta^{HN}(G)$ 
plays the key role among subsequent works.
Nowadays, it is widely considered 
that the lower bound $C_\theta^{HN}(G)$ gives the ultimate bound 
in state estimation.

\section{Quantum estimation with group covariant estimator}\Label{S4}
Quantum system has useful group symmetry, which is an essential
difference from classical system.
It is not impossible to consider a group symmetry in a classical system, but
the irreducible decomposition does not have a simple form in a classical system.
In contrast, 
Since a quantum system often has a very elegant irreducible decomposition,
we can expect that group symmetry greatly simplifies the problem for quantum state estimation. 
Based on this idea, Holevo formulated quantum state estimation theory 
when the state family ${\cal S}=\{\rho_{\theta}\}_{\theta \in \Theta}$
has the symmetry with respect to a representation $f$ of a group $G$ 
on a Hilbert space ${\cal H}$ \cite{Hol-Cov,Hol-Pro}.
When the parameter space $\Theta$ is closed with respect to an action of the group $G$ and the state family ${\cal S}$ satisfies the following condition,
the state family is called a covariant family with respect to the representation 
$f$.
\begin{align}
f(g)\rho_\theta f(g)^\dagger = \rho_{g\theta}
\end{align}
for $g\in G$ and $\theta \in \Theta$.
In this section, we assume the above condition for our state family ${\cal S}$. 
Next, we consider a POVM $\Pi$ whose data set is $\Theta$.
We say that a POVM $\Pi$ is covariant with respect to the representation $f$
when the following condition holds:
\begin{align}
f(g) M(B) f(g)^\dagger=M(g B)
\end{align}
for any Borel set $B$ of $\Theta$, where $gB:=\{g \theta | \theta \in B\}$.
The concept for the group covariant measurement can be 
attributed back to Mackey \cite{Mackey}.

To discuss the precision of our estimate, we need to employ 
an error function $R(\theta,\hat{\theta})$.
Then, the average error is given as
\begin{align}
R_\theta(\Pi):= \int_{\Theta} R(\theta,\hat{\theta}) 
\Tr \Pi(d\hat{\theta})\rho_\theta.
\end{align}
When we focus on the average for $\theta$, 
using a prior distribution $\nu$ on $\Theta$, we discuss 
the Bayesian risk 
\begin{align}
R_\nu(\Pi):=\int_{\Theta}R_\theta(\Pi) \nu(d\theta).
\end{align}
When $G$ is a compact group,
the invariant probability measure $\mu$ on $G$ can be defined.
We often employ $\mu$ as our prior distribution.
When we are interested in the worst case $\theta$, 
we discuss the minimax risk 
\begin{align}
R(\Pi):=\max_{\theta \in \Theta}R_\theta(\Pi) .
\end{align}

In this problem setting, it is natural to impose the group invariance 
to the error function $R(\theta,\hat{\theta})$, which is formulated as
\begin{align}
R(g\theta,g\hat{\theta})=R(\theta,\hat{\theta}).\Label{NZR}
\end{align}
When the error function $R$ satisfies \eqref{NZR} and 
the estimator $\Pi$ is covariant,
we have
\begin{align}
R_\theta(\Pi) =
R_{\theta'}(\Pi) =R(\Pi)=R_\nu(\Pi)
\end{align}
for $\theta',\theta \in \Theta$.

When the error function $R$ satisfies \eqref{NZR}
and $G$ is a compact group, 
as a quantum version of Hunt-Stein theorem \cite{Ferguson},
Holevo \cite{Hol-Cov,Hol-Pro} showed 
the following relations.
\begin{align}
\min_{\Pi\in {\cal M}}R(\Pi)=\min_{\Pi\in {\cal M}} R_\mu(\Pi)=
\min_{\Pi \in {\cal M}_{cov}}R(\Pi)
=\min_{\Pi\in {\cal M}_{cov}} R_\mu(\Pi)
=\min_{\Pi\in {\cal M}_{cov}} R_\theta(\Pi),\Label{XAR}
\end{align}
where 
${\cal M}$ expresses the set of POVM whose data set is $\Theta$
and ${\cal M}_{cov}$ expresses the set of covariant POVM whose data set is $\Theta$.

When $G$ is not compact, the invariant prior $\mu$ does not exist.
Instead of \eqref{XAR}, we have \cite{Bogomolov,Ozawa}
\begin{align}
\min_{\Pi\in {\cal M}}R(\Pi)=
\min_{\Pi \in {\cal M}_{cov}}R(\Pi)
=\min_{\Pi\in {\cal M}_{cov}} R_\theta(\Pi).\Label{XAR2}
\end{align}

Now, we assume that there exists an invariant measure $\mu$
on $\Theta$, which is a weaker condition than the compactness of $\Theta$.
For a covariant POVM $\Pi$, there exists 
a positive semi-definite operator 
$T_{0}$ such that
\begin{align}
M(d\theta)= f(g_\theta) T_{0} f(g_\theta)^\dagger \mu(d\theta),\Label{ZBY}
\end{align}
where $g_\theta$ is chosen as $g_\theta \theta_0=\theta$,
and 
\begin{align}
\int_{\Theta} f(g_\theta)T_0 f(g_\theta)^\dagger \mu(d\theta)=I .
\Label{ZAIO}
\end{align}
Conversely, given a positive semi-definite operator 
$T_{0}$ satisfying \eqref{ZAIO}, 
we can construct a covariant POVM $\Pi$ with \eqref{ZBY}.
Therefore, our optimization in \eqref{XAR} and \eqref{XAR2}
is simplified to the optimization for the choice of 
a positive semi-definite operator 
$T_{0}$ satisfying \eqref{ZAIO}.

Furthermore, condition \eqref{ZAIO} can be simplified into 
the case with irreducible representation and 
the case of a commutative group $G$ without multiplicity.
In the case with irreducible representation, 
the left hand side of \eqref{ZAIO} is a constant multiplication of $I$.
Hence, the condition \eqref{ZAIO} is replaced by the normalization of
the trace of $T_0$.
In the case of a commutative compact group $G$ without multiplicity,
each irreducible representation is one-dimensional.
The set of irreducible representations is denoted by $\hat{G}$.
We denote the vector in the irreducible representation space 
identified by $\lambda \in \hat{G}$ by $|e_\lambda \rangle$.
The condition ``without multiplicity'' means that the uniqueness
of the vector $|e_\lambda \rangle$.
We denote the set of irreducible representations appearing in ${\cal H}$
by $\hat{S}$.
Then, any vector in ${\cal H}$ is written as 
$\sum_{\lambda \in \hat{S}} a_\lambda |e_\lambda \rangle$.
The condition \eqref{ZAIO} is simplified as
$ \|a_\lambda\|^2=1$ for $ \lambda \in \hat{S}$.

Holevo \cite{Hol-Cov,Hol-Pro} applied this approach to several covariant models. 
The first example is the model with the group U(1), in which Helstrom \cite{Helstrom74} studied the case when
the error function is the delta function.
In this model, a state $\rho$ and a representation of the group U(1) are fixed.
This model can be considered as 
the estimation of unknown phase operation. 
Holevo derived the optimal estimator under proper error functions.
The second example is a model with Weyl-Heinsenberg group.
He discusses the estimation in quantum Gaussian family \cite[Chapter 5]{Hol-Pro}.
The third example is a model with group SU(2) \cite{Hol-Cov}.
 
Later, this approach diverged into various directions.
The first one is the asymptotic estimation on the full pure state family \cite{Ha98}. 
The i.i.d. extension of the full pure state family is written as 
a state family on the symmetric tensor product space, which is an irreducible space.
Hence, this approach exactly derives the optimal estimator with a finite size.
In contrast, the Cram\'{e}r-Rao approach gives only an asymptotically approximately optimal estimator. 
In this way, while the group covariant approach can be applied limited cases,
this approach can drive a stronger conclusion in this way.

The second one is the estimation of unitary action.
When we estimate the unknown applied unitary and 
the set of possible unitaries forms a group,
the group covariant approach works well.
In this case, we can optimize the input state as well as the measurement.
To consider this problem, many researchers \cite{LP,BDM,CDS,BBM,CDPS2,Ha-PLA,IH09,H-CMP}
used covariant measurements.
Thanks to the covariant state estimation theory,
the problem is simplified to the optimization of the input state.
As results, they could derive Heisenberg scaling in this problem setting.
In this approach, some of them exactly derived the optimal error with the finite size.
Indeed, many researchers \cite{GLM,GLM2,NOOST,OHNOST,JKFABBM} claimed the achievement of Heisenberg scaling
by using the Cram\'{e}r-Rao approach.
However, the Cram\'{e}r-Rao approach has serious drawback 
in the estimation of unitary action.
Basically, Cram\'{e}r-Rao approach realizes the local optimal estimator.
Using the two-step adaptive method \cite{HMada}, this approach realizes 
the global optimal estimator in the case of state estimation.
However, the two-step adaptive method works only with the usual scaling \cite{Ha-CMP2}.
In addition, the bound obtained by the Cram\'{e}r-Rao approach
is smaller than the limit of the minimum error with a finite size \cite{Ha-CMP2}.
This relation shows that 
it is impossible to show the achievability of the bound obtained by
the Cram\'{e}r-Rao approach.
That is, if it is possible, we can derive an asymptotic bound that contradicts with 
the limit of the minimum error with a finite size.
In this way, Holevo's covariant state estimation theory
greatly supports many researches for Heisenberg scaling.

The key idea of Holevo's group symmetric approach
is reducing the number of free parameters by using the symmetry.
Nowadays, this idea is widely accepted in quantum information theory,
and his result is the first successful example in this direction.
For example, approximate cloning employs the group symmetry \cite{Wener}.
That is, there exists this philosophy behind many recent results of quantum information theory with group covariance
\cite{Ha-Group2}.

\section{Conclusion}\Label{S5}
We have reviewed Holevo's researches in quantum information theory in the 20th century.
Although majority of the research were done in the 1970s, they have strongly
influenced our current studies in quantum information theory.
They form the foundations of many topics in quantum information theory. 
That is to say, he contributed many tools for these areas.

Interestingly, he contributed quantum estimation via two different approaches,
the Cram\'{e}ro-Rao approach and the group covariant approach - he addressed.
That is, he addressed quantum estimation without sticking to one approach.
This fact shows that 
he has the ability to tackle key 
problem without being constrained to a specific
strategy.
The author would like to conclude this article 
by expressing his sincere respect for Holevo's significant contributions to quantum information theory.

\end{document}